# Deep-Learned Observation Operators for Artificial Intelligence Weather Forecasting Models

Kelsey Lieberman[a], Laura Slivinski[b], Matt Bender[a], Chris Miller[a], Josh DaRosa[a], Nick Krall[a], Mohammad Ridhwaan Alam[a], Nick Silverman[a], Sergey Frolov[b]

[a] *MITRE*

[b] *NOAA*

## ABSTRACT

Satellite observation operators play an essential role in atmospheric data assimilation by translating model state variables into observation space. Previous work has shown that deep-learned emulators can effectively predict the outputs of classic observation operators, like the Community Radiative Transfer Model (CRTM), with reduced inference time. This study expands previous work to show the potential for integrating observation operators into artificial intelligence (AI) weather forecasting models. Specifically, this study shows that (1) deep-learned models can effectively predict the innovations (or differences between the simulated and observed radiances) used by data assimilation models and (2) deep-learned observation models suffer only minor degradations in performance when the model state is represented with fewer vertical levels, as is commonly used by AI forecasting models. Experiments were performed using the Unified Forecast System (UFS) replay dataset, including Gridpoint Statistical Interpolation (GSI) observational data for the Advanced Technology Microwave Sounder (ATMS) sensor from 2022 and 2023.

## SIGNIFICANCE STATEMENT

This study demonstrates that deep-learned observation operators can emulate radiative transfer calculations and directly predict satellite radiance innovations, enabling the use of an observation operator with AI-based weather forecast systems. Using UFS replay with GSI ATMS data (2022–2023), we show near-CRTM accuracy with faster inference and only minor degradation when the model state is represented on fewer vertical levels, as is common in AI models. These results indicate practical pathways for end-to-end, computationally efficient AI weather prediction that leverages observations more directly, lowers operational cost, and scales to massive observation streams. The approach bridges traditional numerical weather prediction (NWP) and AI forecasting.

## 1. Introduction

Accurate estimation of atmospheric model states (or initial conditions) is imperative for reliable weather forecasts. Atmospheric model states typically include variables such as temperature, humidity, precipitation, and wind represented over a three-dimensional grid of latitude, longitude, and some type of vertical level at one time. Observed data, such as measurements from weather stations and remote sensing instruments, play an important role in improving the estimates of model states. To assimilate these observations, an observation operator (commonly denoted as the $H$ operator) translates information about a model state $x$ to an observation state $y$. In other words, the observation operator $H$ models the function $H(x) = y$.

The focus of this paper is on observational operators that use radiative transfer theory to translate the physical state variables—such as temperature profiles, humidity profiles, and surface boundary information—to brightness temperatures observed by microwave radiometers on satellite platforms. An example of such an observational operator in classical data assimilation is the Community Radiative Transfer Model (CRTM) (Han et al. 2006; Johnson et al. 2023), which performs a table look-up for radiative transfer functions computed for the narrow bands associated with commonly used remote sensors such as the Advanced Technology Microwave Sounder (ATMS) sensor that we study in this paper. Traditional models like CRTM have two main limitations. First, they are computationally expensive, which can result in them being a bottleneck for forecasting models. Second, they have been developed with an assumption that they could be used in tandem with traditional forecast models that provide a rich set of input features such as extensive vertical resolution and detailed representation of cloud and surface properties. In contrast, new AI forecast models (Lam et al. 2023) utilize fewer input levels, have lower model tops, and carry a limited set of surface variables. Prior work has addressed the first limitation by showing that deep learning models can effectively emulate the outputs of the CRTM with a fraction of the computation (Liang et al. 2022; Howard et al. 2025). This paper expands on this work to show that deep-learned models can directly predict *innovations,* defined as observations minus simulated background, which are used by data assimilation models. We focus on the ability to predict innovations because they form the key object in the state estimation theory and are better conditioned (mean of 0 and a typical range on the order of 1-10K) compared to observations (typical range of 100K). Additionally, we develop observation operators for AI forecast model states. Due to computational constraints, AI forecasting models use fewer vertical levels in the model state than reanalysis datasets typically provide. For example, atmospheric model states in the UFS replay dataset have 127 vertical levels (NOAA 2024); however, GraphCast only uses 13 vertical levels (Lam et al. 2023). We analyze how this sparser model state affects the accuracy of predictions by the observation operator and find that this only results in a small degradation in performance.

We experiment with observations from the ATMS sensor (Kim et al. 2014), which flies on the NPP, NOAA-20, and NOAA-21 satellites in a sun-synchronous orbit. ATMS has 22 channels that are sensitive to physical properties of the atmosphere—such as temperature, moisture, and cloud properties—as well as surface temperature and conditions—such as surface water, ice, snow, vegetation, and soil type. The ATMS sensor is part of the microwave constellation with frequencies that can penetrate clouds and that account for most of the positive observational impact on forecast scores (Geer et al. 2017).

## 2. Related Work

To assimilate satellite observations, current data assimilation modules rely on observation operators to simulate the expected observation value given the model state. Then data assimilation models take the difference between this simulated value and the actual observation, also known as the innovation, or the observation minus background, as input. Observation operators for satellite radiances have traditionally relied on physics-based radiative transfer models, such as Radiative Transfer for TOVS (RTTOV), Community Radiative Transfer Model (CRTM) and the Advanced Radiative Transfer Modeling System (ARMS) (Saunders et al. 1999; Han 2006; Yang et al. 2020). While successive iterations of these models have attempted to improve their efficiency, these remain a bottleneck, especially given the proliferation of sensors and available observations.

In recent years, related work has proposed using deep learning models to emulate physics-based models, like CRTM, in a fraction of the time (Liang et al. 2019; Le et al. 2020; Liang et al. 2022; Stegmann et al. 2022; Howard et al. 2025; Li et al. 2025). Le et al. (2020) use neural networks to model hyperspectral radiative transfer models (RTM) quickly. Similarly, Liang and Liu (2020) train a fully connected neural network (FCNN) to emulate the Visible Infrared Imaging Radiometer Suite (VIIRS) sensor and successive work by Liang et al. (2022) trains an FCNN to emulate the CRTM simulations of the ATMS observations on clear-sky ocean points. Ukkonen (2022) trained an FCNN and a recurrent neural network (RNN) to predict atmospheric radiative transfer, including radiative fluxes, gas optics, reflectance and transmittance. More recently, Howard et al. (2025) trained an FCNN to emulate the CRTM simulations of the infrared channels on the GOES-ABI sensor, and Li et al. (2025) trained a deep learning model to emulate the Geostationary Interferometric Infrared Sounder (GIIRS) on clear-sky ocean conditions. This manuscript extends this work by showing how deep-learned observation models fit into end-to-end AI models, which include both data assimilation and forecasting components.

## 3. Method

We train deep-learned observation models using NOAA's Unified Forecast System (UFS) replay dataset (NOAA 2024). UFS replay is produced by nudging (Orbe et al. 2017) the NOAA UFS coupled model (at 1/4 degree) to existing high-quality reanalysis products: ERA5 for the atmosphere (Hersbach et al. 2020) and ORAS5 for the ocean (Zuo et al. 2019). As a result, this dataset provides a realistic coupled trajectory of the Earth system combined with output every 3 hours from 1994 to 2025. Outputs from the UFS replay were processed using the Gridpoint Statistical Interpolation (GSI) data assimilation system (Kleist et al. 2009). GSI, which is part of NOAA's operational Global Data Assimilation System (GDAS), utilizes advanced quality control and the CRTM observation operator to simulate historic conventional and remote atmospheric observations. This paper focuses on the ATMS sensor and uses GSI outputs to experiment with different formulations of a deep-learned observation operator.

*a. Model formulations*

We experiment with three different formulations of a deep-learned observation operator: (1) an operator that predicts CRTM-simulated brightness temperatures (BTs), (2) an operator that predicts the observed BTs, and (3) an operator that predicts the innovations, or the difference between the observed BTs and the CRTM-simulated BTs. In all cases, we extract a vertical column of data from the model state at the latitude and longitude of the observation using nearest-neighbor interpolation and use this information to predict the values associated with the observation. Thus, all models have 1D inputs and 1D outputs.

Before presenting the three formulations, we introduce some notation. Let $x^b$ denote the estimated (background) model state and $y^o$ represent the observed BTs at a particular time. Then $y^b$ is the CRTM-simulated BTs for model state $x^b$ and $y^i = y^o - y^b$ is the innovation. A data assimilation model takes the background model state $x^b$ and the innovation $y^i$ and nudges the model state to obtain the analysis model state $x^a$. Figure 1 shows example values of $y^b, y^o$ and $y^i$ from channel 1 of the ATMS sensor (23.8 GHz). Note that these values were extracted directly from GSI output files and no ML models were used to obtain these.

The first model formulation we experiment with is deep-learned models that emulate the physics-based observation operator, CRTM, and predict $y^b$. This shows the potential for deep-learned models to serve as an efficient proxy for the CRTM and is similar to the previous studies of Liang et al. (2019), Le et al. (2020), Liang et al. (2022), Howard et al. (2025), and Li et al. (2025). For each observation point, we extract features of the model state $x^b$ at the latitude and longitude point of the observation, angles of the satellite for that observation, and GSI-computed emissivity at that point. We then use all of these features to predict the BTs simulated by the CRTM $y^b$. We denote the predictions of these models $\widehat{y^b}$.

The second model formulation is deep-learned models that predict the actual observations $y^o$. While these models are very accurate at predicting actual observations $y^o$, they are trained using a biased model state $x^b$. Thus, they learn the biases between the model state and the observations to *minimize* the innovations, instead of predicting accurate innovations that would inform the data assimilation model of the biases in the model state. To train models with this formulation, we use the same input features as the previous formulation (features of the model state at the observation point and angles of the satellite), but predict the observed BTs $y^o$, instead of the CRTM simulated BTs $y^b$. We denote the predictions of these models $\widehat{y^o}$.

The third model formulation is deep-learned models that directly predict the *innovation* $y^i = y^o - y^b$. This formulation is useful because data assimilation systems typically assimilate innovations $y^i$ instead of observations $y^o$ and if we train models to predict $y^i$ directly, we can add the observed BTs $y^o$ as input to get a more accurate predictions of the innovations. This formulation takes the same features as the first two plus the observed BTs as input. We denote the predictions of these models $\widehat{y^i}$.

Table 1 summarizes the input and output features for each of these model formulations as well as a sparser version of formulation 3, which we discuss in the next section. For all formulations, we use the same set of input features from the surface and atmospheric model states. These include temperature and water vapor profiles, surface temperature and pressure, wind, cloud and hydrometeor variables, and GSI-computed surface emissivity. We also include sensor angles (zenith, scan, and azimuth) as input features. The only difference in input features between the three formulations is that the innovations formulation also includes the actual observed BTs as input features. The output features differ between the three formulations as described above.

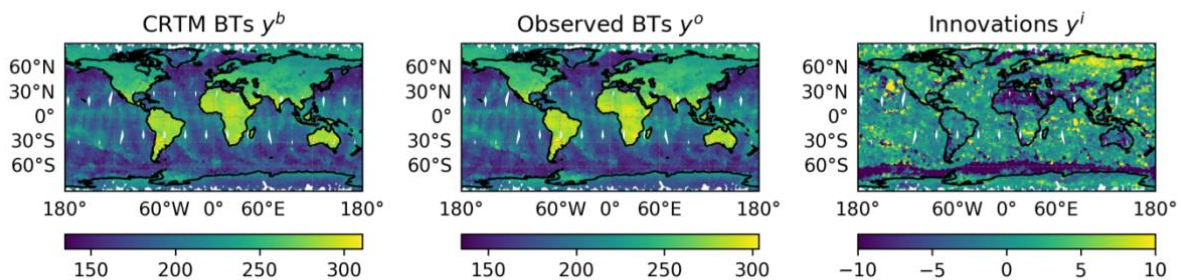

Figure 1: Example of CRTM BTs, observed BTs, and innovations in Kelvin. These observations are for channel 1 of the ATMS sensor (23.8 GHz) on January 4, 2023 and were obtained directly from the UFS replay repository.

|  | Formulation 1: predict $y^b$ (127 model levels) | Formulation 2: predict $y^o$ 127 model levels) | Formulation 3: predict $y^i$ (127 model levels) | Formulation 3: predict $y^i$ (13 model levels) |
|---|---|---|---|---|
| **Input features** | | | | |
| Temperature profile | 127 | 127 | 127 | 13 |
| Water vapor profile | 127 | 127 | 127 | 13 |
| Surface temperature | 1 | 1 | 1 | 1 |
| Surface pressure | 1 | 1 | 1 | 1 |
| Wind | 2 | 2 | 2 | 2 |
| Cloud water mixing ratio | 1 | 1 | 1 | 1 |
| Cloud ice mixing ratio | 1 | 1 | 1 | 1 |
| Rain water mixing ratio | 1 | 1 | 1 | 1 |
| Snow water mixing ratio | 1 | 1 | 1 | 1 |
| Graupel mixing ratio | 1 | 1 | 1 | 1 |
| Sensor zenith angle | 1 | 1 | 1 | 1 |
| Sensor scan angle | 1 | 1 | 1 | 1 |
| Sensor azimuth angle | 1 | 1 | 1 | 1 |
| GSI emissivity | 22 | 22 | 22 | 22 |
| Observed BTs ($y^o$) |  |  | 22 | 22 |
| **Total inputs** | **288** | **288** | **310** | **82** |
| **Output features** | | | | |
| CRTM BTs ($y^b$) | 22 |  |  |  |
| Observed BTs ($y^o$) |  | 22 |  |  |
| Innovations ($y^i$) |  |  | 22 | 22 |
| **Total outputs** | **22** | **22** | **22** | **22** |

Table 1: The number of input and output features used for each deep-learned model formulation.

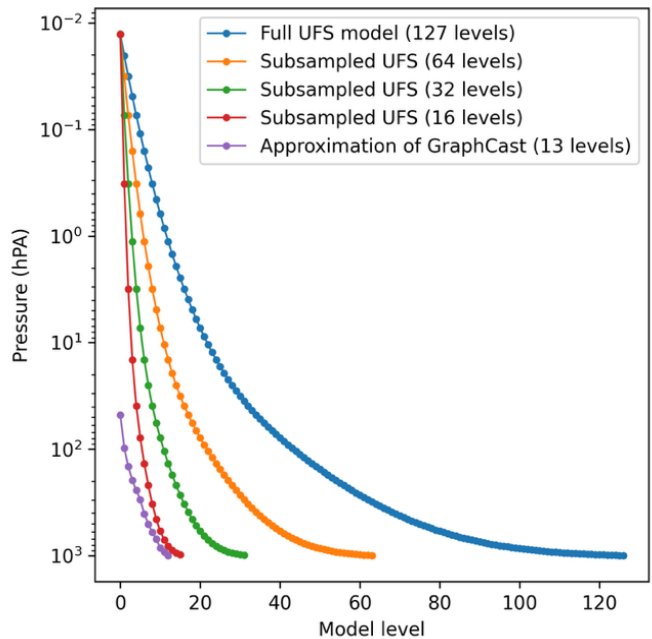

Figure 2: Model level vs. pressure level for model states with sparser vertical levels.

*b. Sparser vertical levels*

The UFS replay dataset has 127 model (vertical) levels, which, while common for physical models, far exceeds the number of levels in AI models (~13-37). Although this level of granularity provides more detailed information about the atmospheric state, it also increases the computational complexity of the AI forecast model. In particular, an AI global forward model must load the full atmospheric state into GPU memory. Thus, it is likely the model state will need to be represented with fewer model levels, and hence the observation operator will need to work in this scenario too. To evaluate the potential for observation models to work with AI forecasting models, we analyze how reducing the number of vertical levels in the model state impacts the performance of observation operators. We train models using features from only 64, 32, or 16 of the UFS replay model levels, instead of all 127, by subsampling every 2nd, 4th, or 8th level. We also try using the same 13 levels from GraphCast (Lam et al. 2023). Figure 2 visualizes the model levels in each of these states over pressure. Table 1 (last column) also shows how using the 13 GraphCast model levels affects the number of input features in the model. Note that for our experiments with 64, 32, and 16 vertical levels, we retain the UFS model top of 1 hPa, while the model top for the GraphCast levels was 50 hPa. For comparison, the UFS model has 35 model levels between above 50 hPa.

*c. Data sources*

We extract all data for experiments from the NOAA UFS replay repository (NOAA 2024) and train models using data sampled every three days throughout 2022. This frequency was chosen to align with the temporal scales of synoptic weather patterns in the mid-latitudes, which are dominated by baroclinic waves that have a life period or a cycle of 3-7 days (e.g., a winter storm cycle usually lasts 3-5 days). By adopting a three-day sampling interval, we ensure comprehensive seasonal converge without oversampling redundant data points from the same weather event. For each day, we used data from all four available times (00, 06, 12, 18 UTC). Surface and atmospheric features came from the 6-hour background forecast fields, while observation and CRTM data came from the GSI diagnostic of the ATMS sensor on the background fields. Each observation sample consists of brightness temperatures for 22 channels at a specific latitude and longitude point. For each observation point, we use the closest vertical column of relevant information about the model state using nearest-neighbor interpolation. We also experimented with using bilinear interpolation and found that this did not significantly affect model performance, so we used nearest-neighbor interpolation for simplicity.

*d. Data scenes: sets of samples*

We experimented with four different types of scenes, or sets of samples, to understand how the model's performance varies depending on the type of data it works on. The four types of scenes that we considered are:

1. **All points:** all observation samples, including land, ocean, and ice points, as well as both clear and cloudy points.
2. **Ocean all-sky points:** the subset of points from the first set that are over the ocean. Since observation points don't fall exactly in the center of the model state grid cell, we define the point to be over the ocean if all its neighboring cells in the model state are classified as being over the ocean. Note that this set does not include ice points.
3. **Clear-sky points**: the subset of points from the first set that are not cloudy, defined by the observed cloud water mixing ratio being ≤ 0.05.
4. **Clear-sky ocean points**: the subset of points from the second set that are not cloudy.

For all sets, we removed any observations that had a missing value in at least one brightness temperature channel (~1.6% of observations). We also partitioned the relevant set of samples into an 80/20 training/testing split and further partitioned the training split into an 80/20 training/validation split. Table 2 shows the number of samples in each set for each of these splits. Note that we only use samples that are included in the GSI files, which is a subset of all available observations for the ATMS sensor.

|  | **Train split (2022)** | **Validation split (2022)** | **Test split (2022)** | **Jan. 4, 2023** |
|---|---|---|---|---|
| All points | 2,989,092 | 747,272 | 934,091 | 41,407 |
| Ocean all-sky points | 1,717,738 | 429,434 | 536,794 | 23,299 |
| Clear-sky points | 2,106,979 | 526,744 | 658,431 | 29,403 |
| Clear-sky ocean points | 835,849 | 208,962 | 261,203 | 11,299 |

Table 2: Number of samples in each scene for training/validation/testing split in 2022 and one representative day, January 4, 2023.

*e. Deep-learned model architecture*

We follow work by Liang et al. (2022) and used a fully connected neural network (FCNN) model architecture with three hidden layers of size 384, 512, and 64. This model also includes three batch normalization layers and ReLU activation. This model has 263k to 353k weights, depending on the number of input features. We train the model using a batch size of 1024, Mean Squared Error (MSE) loss with an L2 regularization penalty of 1e-5, Adam optimization, and a learning rate of 1e-4. We use early stopping with a patience of 20 (i.e., we stop training after the validation loss has not improved for 20 epochs). These configurations were chosen after experimenting with several configurations and finding these to have the most consistent and strong results. Models train for different numbers of epochs depending on their formulation and the scene they work on, with the fastest model (CRTM BTs on the clear-sky ocean scene) finishing in 31 epochs and the slowest model (innovations on the all-sky ocean scene) taking 129 epochs. Code was written with PyTorch and models take 2-10 hours to train on one H100 GPU. For inference, we use the model checkpoint from the epoch with the minimum validation loss.

*f. Feature selection and normalization*

Our input features include the temperature and water vapor at every model level. We also include two cloud hydrometeor variables (cloud water mixing ratio and cloud ice mixing ratio) and thee precipitating hydrometeors (rain water mixing ratio, snow water mixing ratio, and graupel mixing ratio) from the atmospheric profile. Since the mixing ratios tend to be sparse (i.e., they are zero for most of the grid points unless there is a cloud present), we first integrate these variables over the vertical column to compress the information into a two-dimensional field. We integrate these values using the trapezoidal rule over the pressure level, not the model level. We found that vertical integration of cloud properties improved model performance. The input features also include four surface features: temperature, pressure, and two wind variables (u and v components). Finally, we include three angles of the satellite (zenith, scan, and azimuth), and the GSI-computed emissivity for all 22 channels.

Following Liang et al. (2022), we normalize all variables, except for humidity (water vapor) and the hydrometeor mixing ratios. To normalize each scalar feature, we calculate the mean $\mu$ and standard deviation $\sigma$ of the feature in the train set. Then we normalize the original feature values $x$ using $\hat{x} = \frac{x-\mu}{\sigma}$. We normalize both input and output features for training. To get the final predictions, we de-normalize the output features using the inverse process. The water vapor profile is not normalized because humidity is essentially zero above the tropopause, and normalizing near-zero values for the upper levels amplifies numerical noise in the model.

# 4. Results
## a. Results for different model formulations and scenes

Table 3 shows the accuracy of each model formulation at predicting the innovations on the four scenes described in Table 2. For formulations 1 and 2, the innovation is computed by taking the difference between the observations and the model predictions (i.e., $y^o - \widehat{y^b}$ formulation 1 and $y^o - \widehat{y^o}$ for formulation 2). For formulation 3, the innovation is the model prediction (i.e., $\widehat{y^i}$). We measure the performance using the Root Mean Squared Error (RMSE) of the predicted innovations versus the true innovations, averaged over 22 channels. We also visualize the predicted innovations and innovation errors for a test day outside of the training dataset in Figure 3.

The results from formulation 1 (top row of Table 3 and left column of Figure 3) show that deep-learned models can effectively emulate the CRTM model. The accuracy of the model depends on the scene it works on and ranges from an RMSE of 0.913K on clear-sky scenes to 1.369K on ocean all-sky scenes. Since deep-learned models can perform inference faster than classic models, like the CRTM, this suggests that deep learning can be an efficient alternative to classic observation models.

The results from formulation 2 (middle row of Table 3 and middle column of Figure 3) show that while the formulation is the most accurate at predicting the actual observations $y^o$, it is the least accurate at predicting the CRTM innovations. To see this, observe that that this formulation has the smallest magnitude of predicted innovations in Figure 3, but the largest RMSE in Table *3*. This means that the model is learning to correct the biases in the model state $x^b$ during its translation of model state to observation state. While learning this mapping (i.e., minimizing the innovations) could be useful in some contexts, in this case, it results in innovations that are less informative for detecting and correcting biases in the downstream data assimilation step.

Finally, we see that formulation 3, which predicts the innovations directly, has the most accurate predictions of innovations (notice that this formulation has the lowest RMSEs in Table 3 and the smallest differences in Figure 3). This is because it utilizes both the biased model state $x^b$ and the observations $y^o$ to predict the CRTM-simulated innovations. This formulation is the most efficient and effective formulation to find the innovation $y^i$, which can be used by data assimilation systems.

|  | All | Ocean all-sky | Clear-sky | Clear-sky ocean |
|---|---|---|---|---|
| Formulation 1 (predict CRTM BTs $\widehat{y^b}$) $\left\|y^i - (y^o - \widehat{y^b})\right\|_2 = \left\|\widehat{y^b} - y^b\right\|_2$ | 1.214 | 1.369 | 0.913 | 0.915 |
| Formulation 2 (predict observed BTs $\widehat{y^o}$) $\left\|y^i - (y^o - \widehat{y^o})\right\|_2 = \left\|\widehat{y^o} - y^b\right\|_2$ | 2.276 | 1.974 | 2.168 | 1.240 |
| Formulation 3 (predict innovations $\widehat{y^i}$) $\left\|y^i - \widehat{y^i}\right\|_2$ | 0.952 | 1.110 | 0.611 | 0.688 |

Table 3: RMSE between the predicted innovations and the actual innovations for each of the three deep-learned model formulations on four different scenes in Kelvin. Models were trained with 127 model levels and results are shown on the 2022 test set and averaged over all 22 ATMS channels.

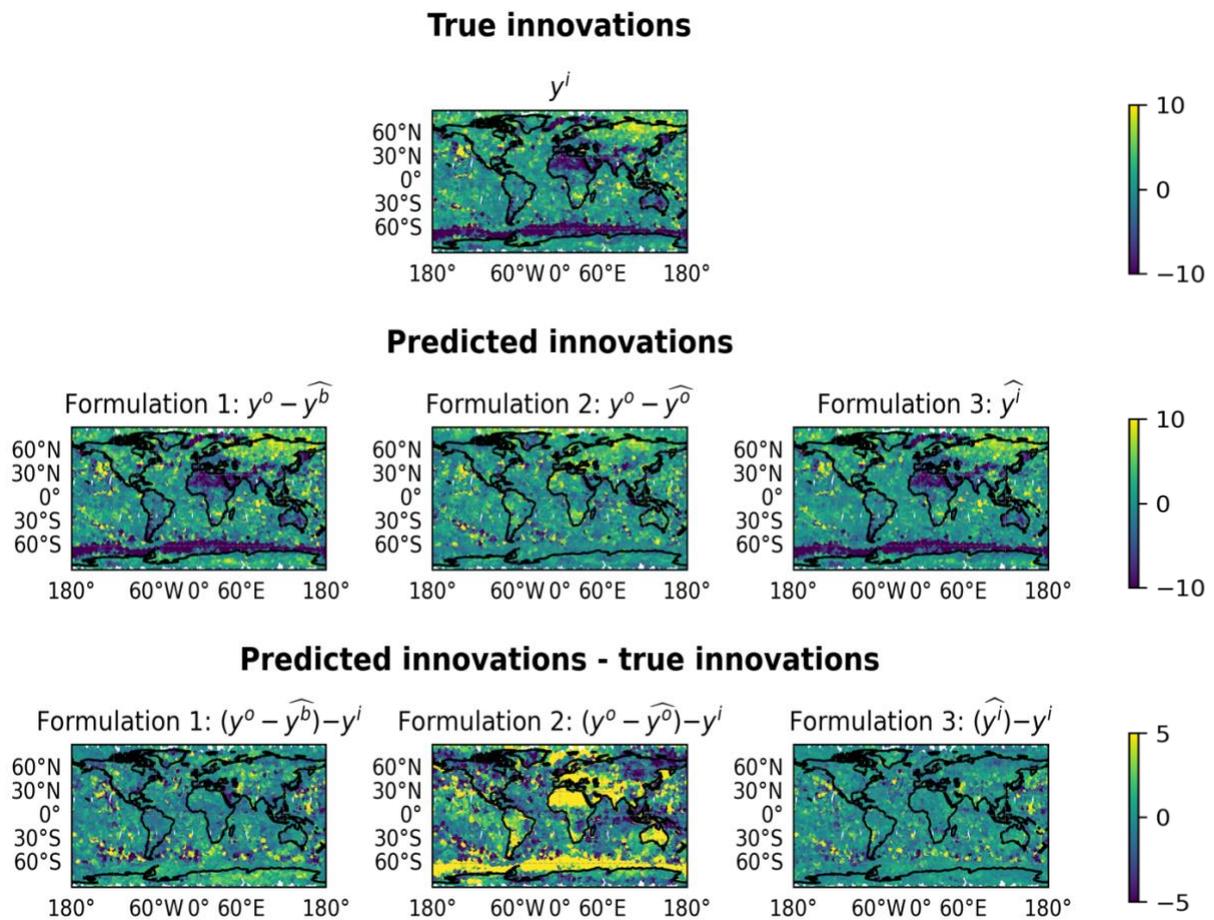

Figure 3: Predicted innovations for each of the three deep-learned model formulations on ATMS channel 1 on January 4, 2023. **Top:** true innovations, **Middle:** predicted innovations, **Bottom:** predicted minus true. Models were trained with 127 model levels on the "all" scene and results are shown in Kelvin.

## b. Results for model states with fewer vertical levels

As discussed in Section 3b, we also experimented with using model states with fewer vertical levels to see how this affects model performance. Figure 4 shows the RMSE on the test set for models that predict innovations using different numbers of vertical levels in the model state. We see that, as expected, the RMSE generally increases as we decrease the number of vertical levels in the model state. However, the increase is relatively small, with the largest increase in RMSE being 5.3%. This suggests that deep-learned observation operators can still perform well even when the model state is represented with fewer vertical levels, which is promising for their integration into AI forecasting models. Note that we do notice some cases where the RMSE decreases as the number of model levels decrease; we find that this is an effect of our normalization procedure and discuss this in more detail in Appendix A.

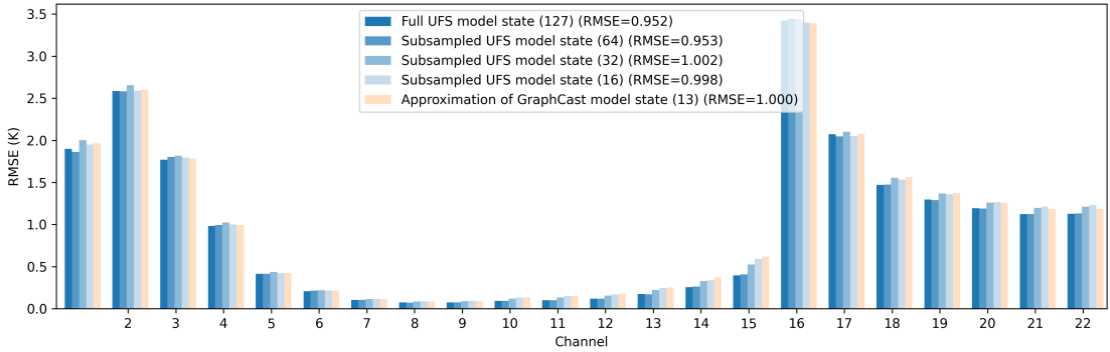

Figure 4: RMSE on the test set for different number of model levels. All results shown are for models trained to predict innovations on the 'all' scene and are in Kelvin. The temperature sounding bands (channels 5-15) have the lowest errors, while the surface sounding bands (channels 1-4 and 16-17) have the highest errors and the humidity sounding bands (channels 18-22) have errors of medium magnitudes.

## c. Ablation studies

To evaluate the relative importance of various input features and model architecture choices, we conduct a series of ablation experiments by systematically removing or replacing key components.

1) EFFECT OF INCLUDING AND INTEGRATING HYDROMETEOR VARIABLES

We found that integrating the five hydrometeor variables (cloud water, cloud ice, rain water, snow water, and graupel) over the vertical profile to create 2D fields improved model performance compared to using the full 3D profiles of these variables. The first, second, and fourth rows of Table 4 show the RMSE when using integrated hydrometeor variables, full-profile hydrometeor variables, and no hydrometeor variables, respectively. We see that using integrated hydrometeor variables results in the lowest RMSE across all scenes, which suggests that integrating these variables reduces the noise and complexity of input features and makes it easier for the model to learn the mapping to brightness temperatures. We also tried adding two concentration features (cloud ice water number concentration and rain number concentration), where concentration is defined as the number of hydrometeor particles per unit volume of air, and show these results in the third row of Table 3. Similar to how mixing ratios are treated, we integrated these features over the pressure level, but, unlike the mixing ratios, we normalized the integrated features. Adding these features slightly degraded performance (RMSE increased by 6%), but this was the second-best configuration overall.

|  | All | Ocean all-sky | Clear-sky | Clear-sky ocean |
|---|---|---|---|---|
| Emissivity + integrated hydrometeors | **0.952** | **1.110** | **0.611** | **0.688** |
| Emissivity + full profile hydrometeors | 0.976 | 1.119 | 0.628 | 0.701 |
| Emissivity + integrated hydrometeors + concentrations | 0.958 | 1.124 | 0.623 | 0.672 |
| Emissivity + no hydrometeors | 1.078 | 1.371 | 0.663 | 0.732 |
| No emissivity + integrated hydrometeors | 1.253 | 1.263 | 1.102 | 0.765 |
| No emissivity + no hydrometeors | 1.394 | 1.360 | 1.111 | 0.820 |
| Precursors for emissivity + integrated hydrometeors | 1.140 | NA | 0.911 | NA |

Table 4: RMSE on the test set for different input features. All results shown are for models trained to predict innovations with 127 model levels and are in Kelvin.

2) EFFECT OF EMISSIVITY

We also trained models without surface emissivity as an input feature to see how this affects model performance. The first and fifth rows of Table 4 show the RMSE when using emissivity and no emissivity, respectively. We see that including emissivity as an input feature significantly improves model performance across all scenes. Additionally, we tried using emissivity precursors, or the features used to compute emissivity, instead of emissivity. We used the following features as precursors: land category, land fraction, vegetation fraction, soil type, soil moisture, ice concentration, surface roughness, snow depth, and snow concentration. The results (last row of Table 4) indicate that adding precursors improves models that lack emissivity but doesn't fully replace it; for example, the all-scene RMSE is 0.952 with emissivity, 1.253 without emissivity, and 1.140 with precursors in lieu of emissivity. Furthermore, on the ocean all-sky and ocean clear-sky scenes, all the precursor features have the same values for all points, which caused training errors; therefore, we exclude results from these scenes. Figure 5 visualizes the effect of emissivity and different hydrometeor setups.

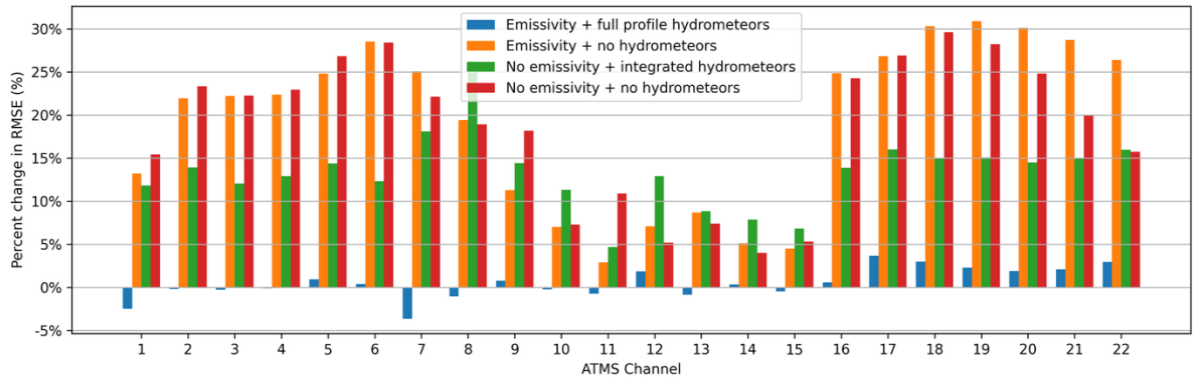

Figure 5: Percent change in RMSE by channel for different input features on the ocean all-sky points. The baseline model used for comparison is the model with emissivity + integrated hydrometeors. Removing hydrometeors causes the RMSE to increase on all channels, with most channels having drastic increases and some temperature bands (channels 10-15) having smaller increases. Removing emissivity causes a relatively consistent increase in RMSE among all bands.

3) EFFECT OF DIFFERENT MODEL ARCHITECTURES

Table 5 shows the RMSE on the test set for different model sizes and architectures. Recall that our baseline model is an FCNN with 3 hidden layers of size: 384, 512, and 64 and batch normalization, which has 350,806 model weights. It also utilizes batch normalization to ensure stable gradients and faster convergence. To explore the relationship between model capacity and error, we experiment with two larger FCNN architectures: a wider 3-layer model (with hidden layers of size: 512, 1024, 62 and 750,144 weights) and deeper 4-layer model (with hidden layers of size: 400, 512, 1024, 64 and 877,574 weights). We also evaluate the impact of the batch normalization to assess whether the internal covariate shift significantly hindered the baseline's performance. The results show that larger models and models without batch normalization have very comparable performance to the baseline model, with slight increases or decreases depending on the scene.

To address potential issues with vanishing gradients in deeper configurations, we also try replacing the FCNN architecture with a residual neural networks (ResNet) architecture (He et al. 2016). ResNet models utilize skip connections to reduce vanishing gradients and

File generated with AMS Word template 2.0

allow for more efficient training. We test two ResNet model architectures with three blocks of two hidden layers each (one with 128 neurons and the other with 256 neurons per layer). Like the other model architecture experiments, we see that all models have comparable results to the baseline. Overall, we see that we can improve the model slightly by increasing the model size, but ultimately changes in the architecture have much smaller and more inconsistent effects than changing the input features.

| | | | RMSE BY SCENE | | | |
|---|---|---|---|---|---|---|
| **MODEL TYPE** | **HIDDEN LAYER STRUCTURE** | **NUMBER OF WEIGHTS IN MODEL** | **ALL** | **ALL-SKY OCEAN** | **CLEAR-SKY** | **CLEAR-SKY OCEAN** |
| FCNN | 3 layers: 384, 512, 64 | 352,726 | 0.952 | 1.110 | 0.611 | 0.688 |
| FCNN | 3 layers: 512, 1024, 64 | 753,344 | **0.935** | **1.102** | 0.615 | 0.668 |
| FCNN | 4 layers: 400, 512, 1024, 64 | 881,446 | 0.939 | 1.110 | **0.580** | 0.671 |
| FCNN w/out batch norm | 3 layers: 384, 512, 64 | 350,806 | 0.947 | 1.121 | 0.587 | 0.662 |
| ResNet | 6 layers: 128 w/ residual connections | 141,718 | 0.980 | 1.144 | 0.609 | 0.667 |
| ResNet | 6 layers: 256 w/ residual connections | 480,022 | 0.982 | 1.146 | 0.600 | **0.660** |

Table 5: RMSE on the test set for different model sizes and architectures. All results shown are for models trained to predict innovations (formulation 3) with 127 model levels and are in Kelvin.

## 5. Conclusion

In this paper, we investigated how ML-based observation operators can effectively replace traditional satellite observational operators based on radiative transfer modeling (such as the CRTM model used to generate targets in this study). Specifically, we asked (1) can we replicate previous deep learning results using the UFS replay dataset, and (2) how can we adapt previous ML models to be integrated into larger AI models that include data assimilation and forecasting.

We found that the ML models developed here produce credible predictions of the model innovations—a critical quantity that informs the data assimilation system about the mismatch between the observed and the modeled atmosphere. Specifically, we found that our ML-based operator was able to accurately predict the innovation quantities even for surface-sensitive and precipitation-sensitive channels. Our results indicate that models trained to predict CRTM-based innovations are the most effective way to predict innovations computed with the classical radiative transfer model. In an end-to-end system, these innovations, along with the forecasted initial conditions, are used as inputs to the data assimilation model, which nudges the initial conditions. These updated initial conditions are then fed into an autoregressive forecast model, like GraphCast (Lam et al., 2023).

We found that the ML-based observation operator had very little degradation (less than 6% increase in RMSE) as we reduced the number of vertical levels for the input features. Specifically, the ML-operator that only used equally spaced 16 subsampled levels (from 1 Pa to the surface) was almost as effective as the operator that used input features on the original 127 levels. Surprisingly, this finding also held when we used the unequally spaced 13 levels from 50 hPa to the surface that are currently used in ML-based forecast models. We also found that we could further collapse the vertically resolved representation of cloud and precipitation hydrometeors into vertically integrated quantities. In fact, the models with vertically integrated quantities were slightly more accurate compared to vertically resolved hydrometeors, possibly due to the lower dimensionality of the former. In future studies, it would be interesting to expand this work to hyperspectral infrared sensors such as the Infrared Atmospheric Sounding Interferometer (IASI) (Blumstein et al. 2004) that can have thousands of highly correlated channels. Could a signal from a sensor with higher spectral resolution be equally as well predicted with just a few vertical levels as was found in this study for the ATMS sensor with 22 spectral channels?

One unexpected finding from this study was the tradeoff between ML models trained to predict CRTM innovations and ML models trained to predict observed brightness temperatures directly. We found that the models trained on the CRTM-based targets (formulations 1 and 3) inherited the systematic errors in the physics-based predictions, specifically over the Sahara and the summer-time polar ice. In both scenes, the systematic errors in CRTM predictions can be attributed to the poor knowledge of the surface properties of the desert sand (Karbou et al. 2006) and the summertime ice (Lawrence et al. 2019). Conversely, the models trained to directly predict observed brightness temperatures removed the systematic errors present in the CRTM-based predictions. One might ask, what is preferable: a more accurate simulation of the observations or a more physically realistic mapping between the imperfect model state and the observation? We suggest that the answer will depend on how the ML model will be used. In cases where the systematic errors in brightness temperature prediction could be attributed to systematic errors of the model state—such as the ice concentrations, vegetation cover, or surface temperature—it might be more beneficial to use an ML model that preserves physical constraints of the radiative transfer model (e.g., formulation 1 or 3). In that case, we could correct the errors in the model state and the forecast model by back-propagating errors from the innovations in the observation space to the space of the model states. However, if the systematic errors can be attributed to the poorly known processes that are not represented in the model state or not modeled by the radiative transfer model, then one might choose to use formulation 2, which removed systematic errors in physical models by directly predicting observed brightness temperatures. An example of the latter could be surface radiative properties related to the age of ice, penetration of the microwave signal within sand, significant radiation scattering from sand and ice, or the presence of melt layers on the ice surface. These tradeoffs and the methods for bias correction have also been discussed in prior work (Janjić et al. 2018, Dee 2005, Auligné et al. 2007).

Finally, we found that inclusion of the hydrometeors and surface emissivities markedly improved prediction of the innovations for all-sky and surface sensitive channels. We found that inclusion of surface emissivity was more impactful (a 24% decrease in RMSE) compared to inclusion of hydrometeors (12% decrease). As mentioned above, we provided hydrometeor information as vertical integrals, and we provided surface emissivity as diagnosed by the GSI data assimilation system from such input quantities as surface roughness, surface wind speed, sea ice concentration, snow water equivalent, surface land cover type, soil moisture, and others. Our deep-learning model that used these precursors to the emissivity calculations instead of the GSI-computed emissivity was less accurate, suggesting room for improvement in future studies.

Our findings suggest that the new class of AI weather forecast models might benefit from the inclusion of the vertically integrated hydrometeors in addition to vertically resolved moisture profiles. We also speculate that it might be beneficial to develop AI forecast models that represent some form of the surface emissivity—either directly or through precursors, such as surface roughness, land use type, or the snow water equivalent.

*Acknowledgments.*

The MITRE team's work was supported by the MITRE Independent Research and Development Program. Sergey Frolov and Laura Slivinski's work was supported by the Inflation Reduction Act funding to the NOAA Physical Sciences Laboratory. We thank Dr. Benjamin Ruston from UCAR/JCSDA for comments and suggestions that contributed to the manuscript. ©2026 The MITRE Corporation. ALL RIGHTS RESERVED.

*Data Availability Statement.*

The data used in this paper is available in the UFS replay repository (NOAA 2024). Code is available at https://github.com/mitre/deep-obs.

APPENDIX A

**Why does sparsification help in some cases?**

In the results, we see that sometimes sparsification improves model performance or decreases the RMSE. Although at first this seems counter-intuitive and a possible indication of overfitting, we explore this further and find that it is a remnant of our post-processing de-normalization step. This section explains this artifact in more detail.

We first note that the subsampled UFS model states are exact subsets of each other. For example, the model state with 16 levels is a subset of the state with 32, and the model state with 32 is a subset of the state with 64. However, the approximation of the GraphCast model state (13 levels) is not an exact subset of the states with 64, 32, or 16 model levels. Thus, we will ignore the GraphCast approximation for now.

When we reduce the number of vertical levels in the model state, we effectively remove input features from the model. We would expect that this would lead to an increase in RMSE because the model has less information to work with. Table A1 shows that while this is the case most of the time, sometimes decreasing the number of model levels results in decreasing the RMSE. For example, in the 'All' scene, the models that work with 16 model levels have lower RMSE (0.998) than those that work with 32 model levels (1.002). Similarly, in the 'Clear-sky' and 'Clear-sky ocean' scenes, the models that work with 64 model levels have lower RMSE than those that work with all 127 model levels.

To understand this anomaly, we consider that the models are not actually trained to optimize the RMSE of the predicted brightness temperatures directly, but rather the RMSE of the *normalized* brightness temperatures, where the normalization is done channel-wise (see Table A3) for the mean and standard deviations used for the normalization). We then look at the RMSE of the normalized outputs in Table A2. Here we see that the RMSE of the normalized outputs always increases as we decrease the number of model levels, which is what we would expect. Thus, the anomalies we see in Table A1 are due to the de-normalization step and not the training itself. Specifically, since each channel has a different standard deviation, it is possible that one model could have a lower RMSE on normalized outputs but a higher RMSE on de-normalized outputs compared to another (see Table A4 for a toy example of this).

We also note that the approximation of the GraphCast model state (13 levels) does perform better than the states with 32 and 16 model levels in the 'Clear-sky' and 'Clear-sky ocean' scenes. This is likely because the GraphCast model state includes level 126, which is near the surface and thus contains important information for predicting brightness temperatures. The subsampled states with 32 and 16 levels do not include this level. Thus, the GraphCast approximation (13 levels) is not directly comparable to the other subsampled states.

|  | All | Ocean all-sky | Clear-sky | Clear-sky ocean |
|---|---|---|---|---|
| Full UFS model state (127) | 0.952 | 1.110 | 0.611 | 0.688 |
| Subsampled UFS model state (64) | 0.953 | 1.118 | 0.602 | 0.676 |
| Subsampled UFS model state (32) | 1.002 | 1.147 | 0.620 | 0.700 |
| Subsampled UFS model state (16) | 0.998 | 1.182 | 0.656 | 0.722 |
| Approximation of GraphCast model state (13) | 1.000 | 1.185 | 0.655 | 0.720 |

Table A1: RMSE of the predicted brightness temperatures on the 2022 test split for different number of model levels and scenes. All results are shown for models trained with the innovations formulation on the 2022 train split. Values are shown in Kelvin.

|  | All | Ocean all-sky | Clear-sky | Clear-sky ocean |
|---|---|---|---|---|
| Full UFS model state (127) | 0.444 | 0.510 | 0.373 | 0.538 |
| Subsampled UFS model state (64) | 0.446 | 0.525 | 0.380 | 0.541 |
| Subsampled UFS model state (32) | 0.507 | 0.577 | 0.425 | 0.599 |
| Subsampled UFS model state (16) | 0.527 | 0.612 | 0.466 | 0.634 |
| Approximation of GraphCast model state (13) | 0.530 | 0.616 | 0.465 | 0.632 |

Table A2: RMSE of the model outputs (normalized brightness temperatures) on the 2022 test split for different number of model levels and scenes. All results are shown for models trained with the innovations formulation on the 2022 train split. Values are shown in normalized units.

| Channel | Mean | Standard deviation | RMSE of model trained w/ normalized outputs before de-normalizing | RMSE of model trained w/ normalized outputs after de-normalizing | RMSE of model trained w/ unnormalized outputs |
|---|---|---|---|---|---|
| Overall | -0.348 | 2.530 | 0.444 | 0.952 | 0.979 |
| 1 | -1.46 | 7.58 | 0.25 | 1.90 | 1.79 |
| 2 | -1.38 | 8.63 | 0.30 | 2.59 | 2.46 |
| 3 | -1.04 | 4.55 | 0.39 | 1.77 | 1.73 |
| 4 | -0.53 | 2.69 | 0.37 | 0.98 | 0.98 |
| 5 | -0.22 | 1.22 | 0.34 | 0.41 | 0.45 |
| 6 | -0.10 | 0.55 | 0.38 | 0.21 | 0.26 |
| 7 | -0.00 | 0.20 | 0.51 | 0.10 | 0.15 |
| 8 | 0.02 | 0.14 | 0.52 | 0.07 | 0.12 |
| 9 | 0.03 | 0.15 | 0.50 | 0.08 | 0.12 |
| 10 | 0.03 | 0.19 | 0.49 | 0.09 | 0.16 |
| 11 | 0.01 | 0.21 | 0.46 | 0.10 | 0.19 |
| 12 | -0.01 | 0.23 | 0.51 | 0.12 | 0.21 |
| 13 | -0.05 | 0.32 | 0.54 | 0.17 | 0.29 |
| 14 | -0.11 | 0.46 | 0.55 | 0.25 | 0.41 |
| 15 | -0.67 | 0.74 | 0.53 | 0.40 | 0.49 |
| 16 | -0.82 | 7.24 | 0.47 | 3.42 | 3.30 |
| 17 | -0.88 | 6.11 | 0.34 | 2.07 | 1.97 |
| 18 | -0.22 | 4.48 | 0.33 | 1.47 | 1.49 |
| 19 | -0.20 | 3.39 | 0.38 | 1.30 | 1.34 |
| 20 | -0.09 | 2.64 | 0.45 | 1.19 | 1.26 |
| 21 | -0.00 | 2.09 | 0.54 | 1.12 | 1.19 |
| 22 | 0.03 | 1.83 | 0.61 | 1.13 | 1.19 |

Table A3: Statistics used for normalization and effect of normalization on RMSE. All results shown are for models that predict innovations using 127 model levels.

|  | Model A | Model B |
|---|---|---|
| Channel 1 RMSE (normalized) | 0.4 | 0.38 |
| Channel 2 RMSE (normalized) | 0.4 | 0.45 |
| Overall RMSE (normalized) | 0.4 | 0.415 |
| Channel 1 RMSE (de-normalized) | 5*0.4=2 | 5*0.38=1.9 |
| Channel 2 RMSE (de-normalized) | 1*0.4=0.4 | 1*0.45=0.45 |
| Overall RMSE (de-normalized) | 1.2 | 1.175 |

Table A4: Toy example of two models where model B has higher RMSE on normalized outputs but lower RMSE on de-normalized outputs compared to model A. Channel 1 has a standard deviation of 5 and channel 2 has a standard deviation of 1. Then a small increase in normalized RMSE (0.415 vs. 0.4) can still result in a decrease in de-normalized RMSE (1.1.75 vs. 1.2) if the increase was driven by a channel with a smaller standard deviation (channel 2 in this case). The overall RMSE is computed as the average RMSE across both channels.

**Effect of training with normalized outputs**

Given the effects of the de-normalizing step that we saw earlier in this section, we also explore training models with unnormalized outputs. Table A5 shows the effect of this. We see that training with normalized outputs consistently outperforms training with unnormalized outputs across all scenes, although the difference is small. Figure A1 shows how these two models compare in terms of RMSE of each channel. As expected, training with unnormalized results leads to more even errors across channels. We see that the model trained on the unnormalized outputs has slightly lower RMSE on the more challenging channels (e.g., channels 1-3 and 16-17), but higher RMSE on the easier channels (e.g., channels 5-15). Overall the models trained with normalized outputs perform better.

|  | All | Ocean all-sky | Clear-sky | Clear-sky ocean |
|---|---|---|---|---|
| Train w/ normalized outputs | **0.952** | **1.110** | **0.611** | **0.688** |
| Train w/ unnormalized outputs | 0.979 | 1.178 | 0.613 | 0.701 |

Table A5: RMSE on the test set with and without normalizing outputs. All results shown are for models trained to predict innovations with 127 model levels and are in Kelvin.

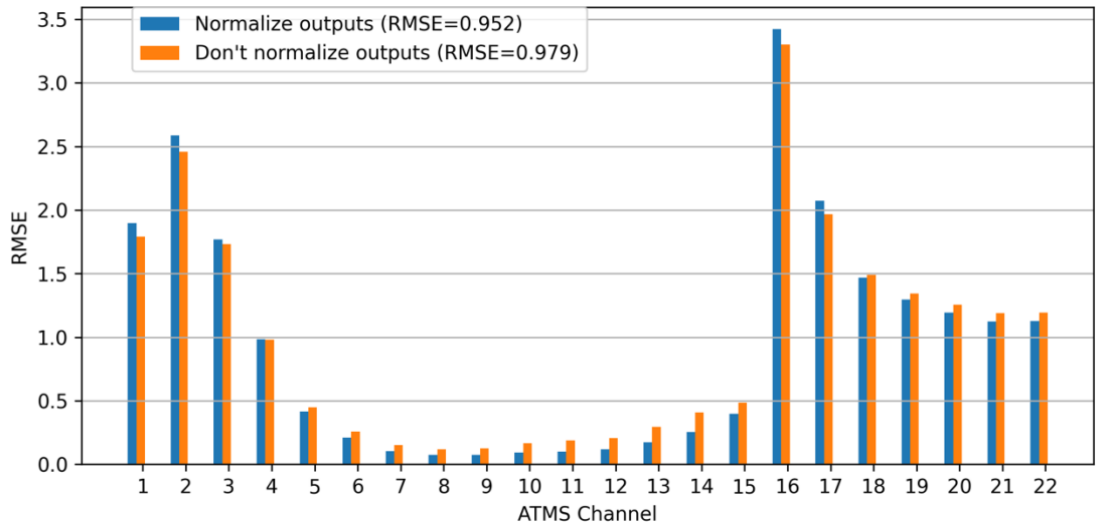

Figure A1: Results on test set when training with normalized and unnormalized outputs on the set of all points.

# REFERENCES

Auligné, T., McNally, A. P., & Dee, D. P. (2007). Adaptive bias correction for satellite data in a numerical weather prediction system. *Quarterly Journal of the Royal Meteorological Society, 133*(624 PART A), 631-642. https://doi.org/10.1002/qj.56.

Blumstein, D., Chalon, G., Carlier, T., Buil, C., Hebert, P., Maciaszek, T., Ponce, G., Phulpin, T., Tournier, B., Simeoni, D., Astruc, P., Clauss, A., Kayal, G., and Jegou, R. (2004). "IASI instrument: technical overview and measured performances", *Proc. SPIE 5543, Infrared Spaceborne Remote Sensing XII*; https://doi.org/10.1117/12.560907.

Dee, D. (2005). "Bias and data assimilation." *Quarterly Journal of the Royal Meteorological Society.*

Geer, A., Ahlgrimm, M., Bechtold, P., Bonavita, M., Bormann, N., English, S., Fielding, M., Forbes, R., Hogan, R., Holm, E., Janiskova, M., Lonitz, K., Lopez, P., Matriicardi, M., Sandu, I., & Weston, P. (2017). Assimilating observations sensitive to cloud and precipitation. *ECMWF Technical Memoranda*. 815.

Han, Y. (2006). *JCSDA Community Radiative Transfer Model (CRTM): Version 1*. https://repository.library.noaa.gov

He, Kaiming; Zhang, Xiangyu; Ren, Shaoqing; Sun, Jian (2016). *Deep Residual Learning for Image Recognition*. Conference on Computer Vision and Pattern Recognition. arXiv:1512.03385. doi:10.1109/CVPR.2016.90.

Hersbach, H., Bell, B., Berrisford, P., et al. (2020). The ERA5 global reanalysis. *Quarterly Journal of the Royal Meteorological Society.* 146: 1999–2049. https://doi.org/10.1002/qj.3803

Howard, L., Subramanian, A. C., Thompson, G., Johnson, B., & Auligne, T. (2025). Probabilistic emulation of the Community Radiative Transfer Model using machine learning. In *arXiv [physics.ao-ph]*. http://arxiv.org/abs/2504.16192

Janjić, T., Bormann, N., Bocquet, M., Carton, J. A, Cohn, S. E., Dance, S. L., Losa, S. N., Nichols, N. K., Potthast, R., Waller, J. A., Weston, P. (2018). On the representation error in data assimilation. *Quarterly Journal of the Royal Meteorological Society.*

Johnson, Benjamin T., Cheng Dang, Patrick Stegmann, Quanhua Liu, Isaac Moradi, and Thomas Auligne. (2023). "The Community Radiative Transfer Model (CRTM): Community-focused collaborative model development accelerating research to operations." *Bulletin of the American Meteorological Society* 104, no. 10: E1817-E1830.

Karbou, F., Gérard, E., & Rabier, F. (2006). Microwave land emissivity and skin temperature for AMSU-A and -B assimilation over land. Quarterly Journal of the Royal Meteorological Society, 132(620), 2333-2355. 10.1256/qj.05.216.

Kim, E., C.-H. J. Lyu, K. Anderson, R. V. Leslie, and W. J. Blackwell (2014), S-NPP ATMS instrument prelaunch and on-orbit performance evaluation, *J. Geophys. Res. Atmos.*, 119, 5653–5670, doi:10.1002/2013JD020483.

Kleist, D. T., Parrish, D. F., Derber, J. C., Treadon, R., Wu, W.-S., & Lord, S. (2009). Introduction of the GSI into the NCEP Global Data Assimilation System. *Weather and Forecasting*, *24*(6), 1691–1705. https://doi.org/10.1175/2009WAF2222201.1

Lam, R., Sanchez-Gonzalez, A., Willson, M., Wirnsberger, P., Fortunato, M., Alet, F., Ravuri, S., Ewalds, T., Eaton-Rosen, Z., Hu, W., Merose, A., Hoyer, S., Holland, G., Vinyals, O., Stott, J., Pritzel, A., Mohamed, S., & Battaglia, P. (2023). *GraphCast: Learning skillful medium-range global weather forecasting* (No. arXiv:2212.12794). arXiv. https://doi.org/10.48550/arXiv.2212.12794.

Lawrence, H., Bormann, N., Geer, A. J., Lu, Q., & English, S. J. (2019). Assessment of the forecast impact of surface-sensitive microwave radiances over land and sea-ice. ECMWF Technical Memorandum, No. 841.

Le, T., Liu, C., Yao, B., Natraj, V., & Yung, Y. L. (2020). Application of machine learning to hyperspectral radiative transfer simulations. *Journal of Quantitative Spectroscopy and Radiative Transfer*, *246*, 106928. https://doi.org/10.1016/j.jqsrt.2020.106928

Li, Y., Han, W., Duan, W., Li, Z., & Li, H. (2025). A Machine Learning-Based Observation Operator for FY-4B GIIRS Brightness Temperatures Considering the Uncertainty of Label Data. *Journal of Geophysical Research: Machine Learning and Computation*, *2*(1), e2024JH000449. https://doi.org/10.1029/2024JH000449

Liang, X., Garrett, K., Liu, Q., Maddy, E. S., Ide, K., & Boukabara, S. (2022). A deep-learning-based microwave radiative transfer emulator for data assimilation and remote sensing. *IEEE J. Sel. Top. Appl. Earth Obs. Remote Sens.*, *15*, 8819–8833. https://doi.org/10.1109/jstars.2022.3210491

Liang, X., & Liu, Q. (2020). Applying Deep Learning to Clear-Sky Radiance Simulation for VIIRS with Community Radiative Transfer Model—Part 2: Model Architecture and Assessment. *Remote Sensing*, *12*(22), 3825. https://doi.org/10.3390/rs12223825

Liang, X., Liu, Q., Yan, B., & Sun, N. (2019). A deep learning trained clear-sky mask algorithm for VIIRS radiometric bias assessment. *Remote Sens. (Basel)*, *12*(1), 78. https://doi.org/10.3390/rs12010078

NOAA. (2024). The global ensemble forecast system (version 13) replay dataset. *NOAA Open Data Dissemination Program*., https://psl.noaa.gov/data/ufs replay/.

Orbe, C., Oman, L. D., Strahan, S. E., Waugh, D. W., Pawson, S., Takacs, L. L., & Molod, A. M. (2017). Large-scale atmospheric transport in GEOS replay simulations. Journal of Advances in Modeling Earth Systems, 9, 2545–2560. https://doi.org/10.1002/2017MS001053

Saunders, R., Matricardi, M., & Brunel, P. (1999). An improved fast radiative transfer model for assimilation of satellite radiance observations. *Quarterly Journal of the Royal Meteorological Society*, *125*(556), 1407–1425. https://doi.org/10.1002/qj.1999.49712555615

Stegmann, P. G., Johnson, B., Moradi, I., Karpowicz, B., & McCarty, W. (2022). A deep learning approach to fast radiative transfer. *Journal of Quantitative Spectroscopy and Radiative Transfer*, *280*, 108088. https://doi.org/10.1016/j.jqsrt.2022.108088

Ukkonen, P. (2022). Exploring Pathways to More Accurate Machine Learning Emulation of Atmospheric Radiative Transfer. *Journal of Advances in Modeling Earth Systems*, *14*(4), e2021MS002875. https://doi.org/10.1029/2021MS002875

Yang, J., Ding, S., Dong, P., Bi, L., & Yi, B. (2020). Advanced radiative transfer modeling system developed for satellite data assimilation and remote sensing applications. *Journal of Quantitative Spectroscopy and Radiative Transfer*, *251*, 107043. https://doi.org/10.1016/j.jqsrt.2020.107043

Zuo, H., Balmaseda, M. A., Tietsche, S., Mogensen, K., and Mayer, M.: The ECMWF operational ensemble reanalysis–analysis system for ocean and sea ice: a description of the system and assessment (2019). *Ocean Sci.,* 15, 779–808, https://doi.org/10.5194/os-15-779-2019